\documentclass[journal]{IEEEtran}
\ifCLASSINFOpdf
\else
\fi
\usepackage{graphics}
\usepackage{amssymb}
\usepackage{amsmath}
\usepackage[dvips]{graphicx}
\usepackage{graphicx}
\usepackage{float}
\floatstyle{ruled}
\usepackage{framed}
\usepackage{subfigure}
\usepackage{tikz}
\newfloat{algorithm}{htbp}{loa}
\floatname{algorithm}{Algorithm}
\begin{document}

\title{Complexity Analysis of CSMA Scheduling via Dependencies Matrix}

\author{\IEEEauthorblockN{Mahdi Azarafrooz and R. Chandramouli}\\
\IEEEauthorblockA{Depatment of Electrical and Computer Engineering \\Stevens Institute of Technology}
}

\maketitle

\begin{abstract}

The complexity of a CSMA algorithm has been translated to the norm properties of a dependencies matrix. 
 The maximum throughput optimization is reformulated by including the dependencies matrix in the formulations. It has been shown that for the interference graphs $\mathcal{G}$ that have minimum vertex cover size $\mathcal{C}(\mathcal{G})=\log n$ where $n$ is the number of the links, the optimal strategy of the links is to transmit with the probability 1, i.e a service-rate agnostic approach.

Several numerical analyses have been conducted in order to illustrate the effect of the interference graph, transmission strategy and arrival rate on the dependencies matrix. 

\end{abstract}

\section{introduction}
``Complexity'', once an ordinary noun describing objects with many interconnected parts, now designates a specific field with so many branches. In this paper a system is considered as complex when it shows \emph{emergence} properties. Emergence in this case refers to a situation where the aggregate of interactions exhibit properties not attained by summation (the whole is more than the sum of its parts). From a design perspective, complex systems should be decomposed into weakly interacting subsystems to avoid such properties. The focus of this paper is on the complexity of scheduling in communication networks.

 The idea of layering for complexity decomposition has been applied previously to the communication network protocols [1].  Although the layering techniques have provided a very efficient platform for communication networks, the arrival of cognition in modern radios has increased the complexity. These cognitive abilities shift the underlying models of communication system from complex physical systems to complex adaptive systems.
This is because of the ability of cognitive nodes to interact with each other in a distributed way, where each node not only learns from the radio environment but also interacts with other nodes. 
The idea of decomposition is a good solution to the situations when there is some coupling or interaction between networking problems. The general idea of decomposition is to break the problem into smaller
ones and solving each of the smaller ones in a distributive manner  [2]. In this work a resource scheduling situation is described where distributed optimization is not efficient due to the emergence properties of the system. This is because the optimization of the whole system is more than the sum of its distributed optimization parts.

The focus of this paper is on carrier sense multiple access (CSMA) scenarios. This is due to its connection to the Markov chain system as the few mathematically describable models for the study of the complex systems. We show that if the scheduling parameters in the CSMA scheduling exceed a specific threshold, the local observations of the links may not be effective for a distributed learning mechanism. This is because the local observation of different links get tied up together in a level that the distant link parameters should be considered to achieve the required efficiency. The question we address here is how arrival rate, interference graph and the simple gradient methods for adjusting the transmission rates, affect the complexity of maximum throughput optimization? We will answer this question by introducing a dependencies matrix into the maximal throughput optimization in the CSMA scheduling. Beside studying the complexity of the CSMA, a direct result of our work is to prove that when the minimum vertex cover of the interference graph is logarithmic in terms of the size of vertexes $O(\log n)$, the suitable strategy for solving the optimization problem is to transmit with probability 1, i.e a service-rate agnostic approach. 

\textbf{Related works} 
It is known that the problem of maximum throughput in a CSMA scheduling is the problem of finding the maximum independent set of the wireless interference graph. Using this intuition a Glauber dynamic \footnote{ Glauber dynamic is a Markov chain monte carlo method that can be used to sample the independent set of a graph according to a product distribution} has been applied to the CSMA problem known as PGD-CMSA (Parallel Glauber Dynamic CSMA) in [3]. We consider a non-parallel version of that work (GD-CSMA) for the ease of modelling. It is proved in [3] that there is no complexity emergence (low mixing time in their context) for complete graphs but we show this is also true for the graphs with the minimum vertex cover size of $\log n$. 

Our problem formulation can be bridged to the design problem of low delay maximal throughput CSMA scenarios [4]. Then the results of our paper can be applied automatically to this sets of problems as well. 
Our work differs from [4] in its optimization formulation. Moreover the focus of this paper is to address the complexity decomposition of distributed learning rather than the low delay scheduling algorithms.  

%[5] states that under a general configuration, there exits no low computational scheme such as a distributed learning mechanism that can achieve low delay maximum throughput. Proposition 1 and 2 in our work demonstrates specific cases under which the low delay (low complexity) design and maximum throughput design are completely decoupled. 

The Markov chain of our studied CSMA is also similar to [5]. Using state decomposition, the authors of [5] provide a constraint on the size of the independent sets of the graph that can guarantee the fast mixing condition of Markov chain. In our work the fast mixing condition is part of the throughput optimization problem, a formulation that has not been addressed to the best of our knowledge.

In order to prove Theorem 5, the gradient descent algorithm similar to [6-7] is used. 

\section{Problem formulation}

Consider an wireless interference graph $G=(E,V)$ with set of $V, |V|=n$ nodes as links and a set of $E$ edges. There is an edge between nodes $v_i$ and $v_j$ if they cannot transmit simultaneously. Let's show a feasible schedule $\textbf{X}$ by a vector of the form $(x_i)_{i\in V}$, with $x_i\in\{0,1\}$ for all $i\in V$. A link $i$ is included in the schedule $\textbf{X}$ if $x_i=1$. $\textbf{X}$ is a feasible schedule if $x_i+x_j\leq 1, \hspace{.1cm} \forall (i,j)\in E$ that is an independent set of interference graph $G$. Let $\Omega\in\{0,1\}^{|V|}$ be the set of all feasible schedules or independent sets of $G$. 
Assume the GD-CSMA scheduling algorithm to be as the following:

For time $t$,
\begin{itemize}
	\item Phase 1: Select a link $i$ uniformly at random.
\end{itemize}

\begin{itemize}
 	\item Phase 2: 
\hspace{.5cm}if $\sum_{j \in \mathcal{N}_i} x_j(t-1)=0$

\hspace{1cm}(a) $x_i(t)=1$ with probability $\mathcal{U}_i=\frac{\lambda_i}{1+\lambda_i}$

\hspace{1cm}(b) $x_i(t)=0$ with probability $1-\mathcal{U}_i=\frac{1}{1+\lambda_i}$

\hspace{.5cm}Else:

\hspace{1cm}$x_i(t)=0$.
	
For every link $j\neq i$:
	
\hspace{.5cm} $x_j(t)=x_j(t-1)$.

\end{itemize} 
where we call $\mathcal{U}_i$ the transmission strategy and $\lambda_i>0$ the fugacity parameter. 

Let the packet arrival distribution of links follows an i.i.d Bernouli distribution with the expected arrival vector of the $\boldsymbol \nu=(\nu_{i}), \forall i$. 
Also let define the capacity region of the network as:
%the set of all arrival rates $\boldsymbol\nu$ for which there exits a scheduling algorithm that can bound the queue length in some appropriate stochastic sense. It is known that the capacity region is defined in \eqref{capdef}:
\begin{eqnarray}\label{capdef}
\begin{array}{l}
\begin{split} 
\Lambda=\{\boldsymbol \nu \geq 0|\exists \boldsymbol \mu \in \text{Co}(\Omega), \boldsymbol \nu < \boldsymbol \mu \}
\end{split} 
\end{array}
\end{eqnarray} 
where $ \text{Co}(\Omega)$ is the convex hull of the set of feasible schedules, i.e, $\boldsymbol \mu \in  \text{Co}(\Omega)$ if $\boldsymbol \mu=\sum_{\textbf{X}\in \Omega} t_{\textbf{X}}\textbf{X}$, where $\sum_{\textbf{X}}t_{\textbf{X}}=1$ and $t_{\textbf{X}}\geq 0$ can be viewed as the fraction of time that schedule $\textbf{X}$ is used. 

%Note that inequalities are component wise. Also let $\Lambda:=\text{Co}(\Omega)$.
The following theorem and the optimization formulation are the direct results of [3].

\textbf{Theorem 1} [3]: 
%Given any $\boldsymbol \nu \in \Lambda$, there exists suitable $\boldsymbol{\lambda}$ such that for every link $i$, its mean service rate $s_i:=\sum_{\textbf{X}:i \in X}\pi(\textbf{X})$ is equal to the mean arrival rate $\nu_i$ under PGD-CSMA with transmission aggregation $\boldsymbol{\lambda}$, where $\pi(\textbf{X})$ is given in \eqref{stationary}. In other words, the service rates of PGD-CSMA can exactly meet the arrival rates of all links.
The dynamics of the GD-CSMA results to that of a Markov chain with the following product-form stationary distribution:
\begin{eqnarray}\label{stationary}
\begin{array}{l}
\begin{split} 
\pi(\textbf{X})=\frac{\prod_{i\in \textbf{X}}\lambda_i}{\sum_{\textbf{X}'\in\Omega}\prod_{i\in \textbf{X}'}\lambda_i}
\end{split} 
\end{array}
\end{eqnarray}

\begin{framed}

Optimization formulation:\\\noindent
Let denote $r_i:=\log(\lambda_i)$. Then given any $\boldsymbol \nu \in \Lambda$, the service rates of the GD-CSMA can exactly meet the arrival rates of all links when the vector $\textbf{r}^{*}=(r_i^{*}),\forall i$ is the solution of the convex optimization problem $\max_{\textbf{r}}F(\textbf{r};\boldsymbol{\nu})$ 
\begin{eqnarray}\label{prime}
\begin{array}{l}
\begin{split} 
F(\textbf{r};\boldsymbol{\nu})=\sum_{i}\nu_ir_i-\log(\sum_{\textbf{X}\in \Omega}\exp(\sum_{i}x_ir_i)).
\\ \text{s.t} \hspace{1cm} r_i\geq 0, \forall i
\end{split} 
\end{array}
\end{eqnarray}

\end{framed}

\subsection{A simple distributed optimization algorithm}
Taking the partial derivative from \eqref{prime} with the substitution for the mean service rate of link $i$ $s_i:=\sum_{\textbf{X}:x_i=1}\pi(\textbf{X})$ yields:
\begin{eqnarray}\label{prime derivitive}
\begin{array}{l}
\begin{split} 
\partial F(\textbf{r},\boldsymbol{\nu}) / \partial r_i=\nu_i-s_i(\textbf{r})
\end{split} 
\end{array}
\end{eqnarray}
Using \eqref{prime derivitive} a simple gradient algorithm of \eqref{alg1} can be suggested. \eqref{alg1} can be perceived as a distributed algorithm since link parameters can be adjusted based on the local informations of arrival rate $\nu_i^{'}(t)$ and service rate $s_i^{'}(\textbf{r}(t))$ as the average arrival rate and service rate between time $t$ and $t+1$.
\begin{eqnarray}\label{alg1}
\begin{array}{l}
\begin{split} 
r_i(t+1)=[r_i(t)+\alpha(\nu_i^{'}(t)-s_i^{'}(\textbf{r}(t)))]_{+}
\end{split}
\end{array}
\end{eqnarray}

\subsubsection{Complexity of the distributed optimization}

The previous distributed approach is feasible only when the average service rate $s_i^{'}(\textbf{r}(t))$ perceived by the link $i$ can track the stationary distribution of the CSMA Markov chain $s_i(\textbf{r})$ fast enough. Let's say it is fast enough when for every link $i$, \eqref{markdist} is bounded above by some polynomial function $O(\text{poly}(n))$:
\begin{eqnarray}\label{markdist}
\begin{array}{l}
\begin{split} 
|\frac{1}{T}\sum_{k=t}^{t+T}s'_i(\textbf{r}(k))-s_i(\textbf{r})|
\end{split} 
\end{array}
\end{eqnarray}  
by remembering that $s_i=\sum_{\textbf{X}:x_i=1}\pi(\textbf{X})$, \eqref{markdist} can be written as:
\begin{eqnarray}\label{markdistG}
\begin{array}{l}
\begin{split} 
|\frac{1}{T}\sum_{k=1}^{T}(\sum_{\textbf{X}\in \Omega: x_i=1}\mu_{\textbf{X}(t),k}(\textbf{X})-\sum_{\textbf{X}:x_i=1}\pi(\textbf{X}))|
\end{split}
\end{array}
\end{eqnarray} 
where $\mu_{\textbf{X}(t),k}(\textbf{X})$ is the distribution of the Markov chain of the schedules after $k$ slots if the Markov chain starts with $\textbf{X}(t)$. The expression \eqref{markdistG} can be understood as the mixing time of the Markov chain $\frac{1}{T}\sum_{k=1}^{T}\|\mu_{\textbf{X}(t),k}-\pi\|_{\text{var}}$ known to be bounded below by a exponentially large function in the numbers of links $n$ for some range of parameter $\textbf{r}$ [8]. 
This means there exits transmission strategy $\boldsymbol{\mathcal{U}}=(\mathcal{U}_i), \forall i$ that the average service rate cannot follow the stationary distribution fast enough. Therefore selecting the parameter $s_i^{'}$ as the reference to update the optimization strategy is ineffective. In other words the individual optimization solutions are coupled with the optimization of other links to a level that the problem cannot be solved distributively. 

Therefore our aim is to include the fast mixing condition in \eqref{prime}. The following section introduces the fast mixing condition as a new constraint in the previous optimization set ups.

In [5] using a state decomposition technique it is shown that if the probability of going to states $\textbf{X}$ corresponding with the independent sets of size more than $\llcorner \frac{n}{2(\Delta-1)}\lrcorner$ with $\Delta$ being the maximum degree of the interference graph is rare and $n$ being the number of nodes (links) then the Markov chain is fast mixing regardless of the Glauber dynamic parameters.
 
In the following section instead we address the optimization problem of maximum throughput under the constraint of fast mixing.

\section{Dependencies matrix as a new constraint of the optimization problem}

\subsection{Preliminaries}

If $\mu$ and $\nu$ are two probability distributions on $\Omega$, then the total variation distance between $\mu$ and $\nu$ is:

\begin{eqnarray}\label{dtv}
\begin{array}{l} 
\begin{split}
d_{TV}:=\displaystyle \max_{A\subset \Omega }|\mu(A)-\nu(A)|=\frac{1}{2}\displaystyle \sum_{x \in \Omega}|\mu(x)-\nu(x)|
\end{split}
\end{array}
\end{eqnarray}

 A coupling between two probability distributions $\mu$ and $\nu$ is a pair of random variables $(X,Y)$ such that
 \begin{itemize}
 \item $(X,Y)$ are defined on a common probability space.
 \item $X$ has distribution $\mu$, and
 \item $Y$ has distribution $\nu$
 \end{itemize}

\textbf{Proposition 1} If $\mu$ and $\nu$ are two probability distributions, then
\begin{eqnarray}\label{prop1}
\begin{array}{l} 
\begin{split}
d_{TV}(\mu,\nu)=\displaystyle \min_{(X,Y) \text{couplings} } \mathcal{P}(X\neq Y).
\end{split}
\end{array}
\end{eqnarray}

\textit{Example.} Let $\Omega=\{0,1\}$ and set $\mu_p(0)=1-p$ and $\mu_{q}(1)=p$ Then
\begin{eqnarray}\label{example}
\begin{array}{l} 
\begin{split}
d_{TV}(\mu,\nu)=\frac{1}{2}(|(1-p)-(1-q)|+|p-q|)=|p-q|
\end{split}
\end{array}
\end{eqnarray}
so the coupling using the uniform variable is optimal.

Let $\textbf{S}_j$ be all the pairs of configuration $(\textbf{X},\textbf{Y})\in \Omega^2 $ agreeing on transmission states of all links except the link $j$. Then the dependencies matrix is defined as $\mathcal{R}:=(R_{ij})$ where $i$ and $j$ are different links and the dependencies of link $j$ on $i$ is:
\begin{eqnarray}\label{distancemetric} 
\begin{array}{l} 
\begin{split} 
R_{ij} = \displaystyle \max_{(\textbf{X},\textbf{Y}) \in \textbf{S}_j} d_{TV}(\mu_i(\textbf{X},.),\mu_i(\textbf{Y},.))
\end{split}
\end{array}
\end{eqnarray} 
where $\mu_i(\textbf{X},.)$ denotes the marginal distribution of the transmission state of link $i$, for configurations sampled from $\pi$ in \eqref{stationary} conditioned on agreeing with $\textbf{X}$ at all other links.

\textbf{Theorem 2: Dobrushin condition} The GD-CSMA has the fast mixing time Markov chain when every row sum of the dependencies matrix $\mathcal{R}$ is less than 1. 
\begin{proof}
Theorem 3 in [10].
\end{proof}
The Dobrushin condition roughly states that there is asymptotically no correlation between the link at a $i$ and the link $j$ with distance $d$ from $i$, as $d$ tends to infinity. [8-9] showed a weaker hypothesis for the Dobrushin condition that requires any operator norm of $\mathcal{R}$ to be less than $1$.

In the next section we include the Dobrushin condition as a constraint in the \emph{dual} optimization of  \eqref{prime}.

\textbf{Theorem 3}
The dual problem of \eqref{prime} can be written as \eqref{dual}
\begin{eqnarray}\label{dual}
\begin{array}{l}
\begin{split} 
\min_{\textbf{X}}\sum_{\textbf{X}\in \Omega}p(\textbf{X})\log(p(\textbf{X}))
\\ \text{s.t} \hspace{1cm} \sum_{\textbf{X}}p(\textbf{X})x_i\geq \nu_i,\forall i \in V\\ \hspace{1.5cm} \sum_{\textbf{X}}p(\textbf{X})=1\\  0 \leq  p(\textbf{X}) \leq 1
\end{split} 
\end{array}
\end{eqnarray}
where $p(\textbf{X})$ is the probability of state $\textbf{X}$ in the CSMA Markov chain. 
\begin{proof}
Refer to Appendix A.
\end{proof}

To include the fast mixing condition in \eqref{dual} let's define an expected dependencies distance matrix $\mathcal{I}$ as the following:
Redefine the dependencies metric \eqref{distancemetric} as:
\begin{eqnarray}\label{distancemetricM} 
\begin{array}{l} 
\begin{split} 
R_{ij}^{\textbf{X}} = \displaystyle d_{TV}(\mu_i(\textbf{X},.),\mu_i(\textbf{Y},.))
\end{split} 
\end{array}
\end{eqnarray} 
where $\textbf{Y}\in \Omega$ is the same as state $\textbf{X}$ in all links except the link $j$. The rest of parameters are the same as \eqref{distancemetric}. Denote the matrix of these parameters with $\mathcal{R}^{\textbf{X}}$. Let's define the expected dependencies matrix $\mathcal{I}$ as:
\begin{eqnarray}\label{expected dependencies}
\begin{array}{l}
\begin{split} 
\mathcal{I}=\sum_{\textbf{X}}p(\textbf{X})\mathcal{R}^{\textbf{X}}
\end{split} 
\end{array}
\end{eqnarray}
%Let $\textbf{X}$ be any configuration. There are only two possibilities for the marginal distribution of $\pi$ at a link $i$, conditioned on the neighbours agreeing with $\textbf{X}$. Either some neighbour is occupied under $\textbf{X}$, in which case $i$ is transmitting with probability 1, or all neighbours silent, in which case $i$ is transmitting with probability $\lambda_i$ and silent otherwise. This means that $\mathcal{R}^{\textbf{X}}=\boldsymbol{\lambda}_\textbf{X} \mathcal{A}$  since the influence of $i$ on $j$ is zero except when $i$ and $j$ are neighbours.  
It is easy to see that the convex combination of probability state transitions and the dependencies matrix keeps the sum of every row $i$ of matrix $\mathcal{I}$ bounded above by $1$ as well. That is:
\begin{eqnarray}\label{distance_condition}
\begin{array}{l}
\begin{split} 
\displaystyle \sum_{\textbf{X}}d_i^{\textbf{X}}p(\textbf{X})<1, \forall i \in V
\end{split} 
\end{array}
\end{eqnarray}
where $d_i^{\textbf{X}}$ is the sum of row $i$ in $\mathcal{R}^{\textbf{X}}$.
Now let's rewrite \eqref{distance_condition} as
\begin{eqnarray}\label{distance_condition1}
\begin{array}{l}
\begin{split} 
\displaystyle \sum_{\textbf{X}}(1-x_i)d_i^{\textbf{X}}p(\textbf{X})+
\displaystyle \sum_{\textbf{X}}x_id_i^{\textbf{X}}p(\textbf{X})<1, \forall i \in V
\end{split} 
\end{array}
\end{eqnarray}
Now using the capacity constraint of \eqref{dual}, 
\begin{eqnarray}\label{distance_condition3}
\begin{array}{l}
\begin{split} 
\displaystyle \sum_{\textbf{X}}(1-x_i)d_i^{\textbf{X}}p(\textbf{X})
<1-d_{i,\min} \nu_i, \forall i \in V
\end{split} 
\end{array}
\end{eqnarray}
where $d_{i,\min}=\min_{\textbf{X}}d_i^{\textbf{X}}$ is the row $i$ sum of the dependencies matrix $\mathcal{R}$. By including \eqref{distance_condition3} in \eqref{dual}, 
we end up with the \eqref{optimation_problem}:
\begin{eqnarray}\label{optimation_problem}
\begin{array}{l}
\begin{split} 
\min_{\textbf{X}} \displaystyle \sum_{\textbf{X}\in \Omega}p(\textbf{X})\log(p(\textbf{X}))
\\ \text{s.t} \hspace{.5cm} \displaystyle \sum_{\textbf{X}}(1-x_i)p(\textbf{X})<\frac{1}{d_{i,\min}}-\nu_i,  \forall i \in V \\ \hspace{1.5cm} \displaystyle \sum_{\textbf{X}}p(\textbf{X})=1\\ \hspace{1.5cm} 0 \leq  p(\textbf{X}) \leq 1 
\end{split} 
\end{array}
\end{eqnarray}
Note that for $d_{i,\min}=1$ the problem is the same as $\eqref{dual}$.
The important difference of \eqref{dual} with  \eqref{optimation_problem} is that $d_{i,\min}$ can be written in terms of the dual variable of this optimization problem as will be shown later. Now that we have embedded the complexity constraint in the dual optimization problem we can now bring back the problem to the prime optimization format by taking the dual of dual as is formulated in Theorem 4. It is important to return to the prime optimization formulation as it provides a proper framework to introduce the graph characteristics into the complexity of distributed optimizations. This is discussed in the Section IV.  

\textbf{Theorem 4}
Dual of the optimization set up \eqref{optimation_problem} is in the form of:
\begin{eqnarray}
\label{optimization_minxing time}
\begin{array}{l}
\begin{split}
 \max_{\boldsymbol \zeta}\mathcal{D}(\boldsymbol \zeta;\boldsymbol \nu)= \\
\sum_{i}(\nu_i-\frac{1-\epsilon}{d_{i,\min}})\zeta_i-\log(\sum_{\textbf{X}\in \Omega}\exp(\sum_{i}(x_i-1)\zeta_i)) 
 \\ \text{s.t} \hspace{1cm} \zeta_i\geq 0, \forall i 
 \end{split} 
\end{array}
\end{eqnarray}
\begin{proof}
Similar to proof of theorem 3.
\end{proof}

\subsection{A service-rate agnostic case }

In this section we first present our main theorem and the rest of the paper till the numerical analyses section is to devoted to prove this Theorem.

\textbf{Theorem 5} For interference graph with the minimum vertex cover size of $O(\log n)$, the update strategy of all links is to transmit with probability 1, i.e independent of the service rate $\textbf{s}=(s_i) \hspace{.1cm} \forall i$ .

\begin{proof}
To prove the previous theorem we use a gradient based algorithm approach similar to [6-7]. The main idea of the following technique is to lower bound the change in the dual value by an auxiliary function and then maximize that bound.

For $\zeta_i>0, \delta>-\zeta_i$,
\begin{eqnarray}
\label{dual bound}
\begin{array}{l}
\begin{split} 
\mathcal{D}(\zeta_i)-\mathcal{D}(\zeta_i+\delta_i)=\\ \log(\sum_{\textbf{X}}\exp(\sum_{i}-\delta_i(1-x_i)))-\delta_i\nu_i+\\(\frac{\zeta_i+\delta_i}{d_{i,\min}(\zeta_i+\delta_i)})-(\frac{\zeta_i}{d_{i,\min}(\zeta_i)})=\\
\log(\sum_{\textbf{X}}\exp(\sum_{i}-\frac{(1-x_i)}{\mathcal{C}}\mathcal{C}\delta_i))+\\
-\delta_i\nu_i+(\frac{\zeta_i+\delta_i}{d_{i,\min}(\zeta_i+\delta_i)})-(\frac{\zeta_i}{d_{i,\min}(\zeta_i)})
\end{split}
\end{array}
\end{eqnarray} 
Select $\mathcal{C}=1+\sum_{i}(1-x_i) \hspace{.1cm} \forall \textbf{X}$. Then by Jensen's inequality
\begin{eqnarray}
\label{jensen inequality}
\begin{array}{l}
\begin{split} 
\log(\sum_{\textbf{X}}\exp(\sum_{i}-\frac{(1-x_i)}{\mathcal{C}}\mathcal{C}\delta_i))<\\ \log(\sum_{\textbf{X}}\sum_{i}\frac{1-x_i}{\mathcal{C}}(\exp(-\mathcal{C}\delta_i))+(1-\frac{1-x_i}{\mathcal{C}}))=\\
\log(1+\sum_{\textbf{X}}\sum_{i}\frac{1-x_i}{\mathcal{C}}(\exp(-\mathcal{C}\delta_i)-1))
\end{split} 
\end{array}
\end{eqnarray}
Using \eqref{jensen inequality}, the \eqref{dual bound} can be written as:
\begin{eqnarray}
\label{dual bound inequality}
\begin{array}{l}
\begin{split}
\mathcal{D}(\zeta_i)-\mathcal{D}(\zeta_i+\delta_i)<\\ \log(\sum_{\textbf{X}}\sum_{i}\frac{1-x_i}{\mathcal{C}}(\exp(-\mathcal{C}\delta_i)-1)) \\ -\delta_i\nu_i+(\frac{\zeta_i+\delta_i}{d_{i,\min}(\zeta_i+\delta_i)})-(\frac{\zeta_i}{d_{i,\min}(\zeta_i)})
\equiv \mathcal{A}(\boldsymbol \zeta,\boldsymbol \delta)
\end{split}
\end{array}
\end{eqnarray}
We use the auxiliary function $ \mathcal{A}(\boldsymbol \zeta,\boldsymbol \delta)$ to design the gradient based algorithm.

To continue the proof we use the lemmas 1 and 2.

%\textbf{Theorem 2} Let as before $\lambda_i$ be the transmission aggregation and let initialize it to $\lambda_i^t \geq 0$.

\textbf{Lemma 1} Under the formulation \eqref{optimization_minxing time} $\exp(\zeta_i)=\frac{1}{\lambda_i}$.
\begin{proof}
It is easy to show that the relation between primary and dual variables are given by:
for every $\textbf{X} \in \Omega$
\begin{eqnarray}
\label{relation}
\begin{array}{l}
\begin{split} 
p(\textbf{X})=\frac{\exp(-\sum_{i}\zeta_i(1-x_i))}{\sum_{\textbf{X}}\exp(-\sum_{i}\zeta_i(1-x_i))}
\end{split}
\end{array}
\end{eqnarray}
comparing with \eqref{stationary} and noting that $(1-x_i)\geq 0$ we can see that $\exp(-\zeta_i)=\lambda_i$ or $\exp(\zeta_i)=\frac{1}{\lambda_i}$.
%Let denote the partial Lagrangian of \eqref{optimation_problem} as $\mathcal{L}(p(\textbf{X}),\boldsymbol \nu)=-H(p(\textbf{X}))+\sum_{i}\zeta_i(\nu_i-\frac{1-\epsilon}{d_{i,\max}})$. Then
%Denote $p^{*}(\textbf{X})=\text{argmax}_{p(\textbf{X})}\mathcal{L}(p(\textbf{X}),\boldsymbol \nu)$. Due to the constraint $\sum_{\textbf{X}}p(\textbf{X})=1$, if we there exits some $w$ such that
%\begin{eqnarray}
%\frac{\partial \mathcal{L}(p(\textbf{X}),\boldsymbol \nu)}{\partial p(x)}=-\log(P^{*}(x))-1+\sum_{i}\zeta_i(1-x_i)
%\end{eqnarray}
\end{proof}
 
\textbf{Lemma 2}
$d_{i,\min}$ can be estimated as $\frac{d_i\lambda_i}{1+\lambda_i}$ where $d_i$ is the graph degree of link $i$. 
\begin{proof}
Let $\textbf{X}$ be any configuration. There are only two possibilities for the marginal distribution of $\pi$ at a link $i$, conditioned on the neighbors agreeing with $\textbf{X}$. Either some neighbor is occupied under $\textbf{X}$, in which case $i$ is transmitting with probability 0, or all neighbor silent, in which case $i$ is transmitting with probability $\frac{\lambda_i}{1+\lambda_i}$ and silent otherwise. This means that $d_{i,\min}$ can be estimated as $\frac{d_i\lambda_i}{1+\lambda_i}$ since the dependencies of $i$ on $j$ is zero except when $i$ and $j$ are neighbors.
\end{proof}

The rest of proof is based on a variational method that is to maximize the \textit{expectation} of the bound $E_{\textbf{X}}(\mathcal{A}(\boldsymbol \zeta,\boldsymbol \delta))$. Using linearity of expectation followed by applying Lemma 1 and 2 and then taking partial derivative of $E_{\textbf{X}}( \mathcal{A}(\boldsymbol \zeta,\boldsymbol \delta))$ yields:
\begin{eqnarray}
\label{derivative aux}
\begin{array}{l}
\begin{split} 
\frac{\partial E_{\textbf{X}}(\mathcal{A}(\boldsymbol \zeta,\boldsymbol \delta))}{\partial \delta_i}=\\ (s_i-1)\exp(-\mathcal{C}\delta_i)-\nu_i+\frac{(1+\zeta_i+\delta_i)\exp(\zeta_i+\delta_i)}{d_i}
\end{split} 
\end{array}
\end{eqnarray}
where $s_i=\sum_{\textbf{X}}p(\textbf{X})x_i$ is the average service rate.
Now if each link $i$ updates according to a gradient algorithm of the following:
\begin{eqnarray}
\label{gradient algorithm}
\begin{array}{l}
\begin{split} 
\zeta_i^{t+1}\leftarrow \max (0,\zeta_i^{t}+\delta^{*}_i(\zeta^{t}))
\end{split} 
\end{array}
\end{eqnarray}
where $\frac{\partial E_{\textbf{X}} \mathcal{A}(\boldsymbol \zeta,\boldsymbol \delta)}{\partial \delta_i}|_{\delta^{*}_i(\zeta^{t})}=0$ then $\max \mathcal{D}(\boldsymbol \zeta;\boldsymbol \nu)$ is achieved.

First note that $\frac{{\partial}^2 E_{\textbf{X}}(\mathcal{A}(\boldsymbol \zeta,\boldsymbol \delta))}{{\partial \delta_i}^2}>0$ and convexity implies that $\delta^{*}_i$ to be found at the corners. To have the gradient algorithm converge, the solution is $\delta^{*}_i=-\boldsymbol \zeta_i$. This requires $\frac{{\partial} E_{\textbf{X}}(\mathcal{A}(\boldsymbol \zeta,\boldsymbol \delta))}{{\partial \delta_i}}<0$.  This condition can be achieved if $\mathcal{C}=O(\log n)$ and $s_i-1 < 0, \forall i$. This is true for all the interference graphs except the complete graph. Since $\zeta_i=0$ means transmission with probability 1 therefore for complete graph there exists a link $i$ that $s_i-1=0$. However for a complete graph $d_i=n$ and the third right hand term of the \eqref{derivative aux} will be zero and again the $\frac{{\partial} E_{\textbf{X}}(\mathcal{A}(\boldsymbol \zeta,\boldsymbol \delta))}{{\partial \delta_i}}=-\nu_i<0$.
 
\begin{figure}

\begin{center}

\begin{tikzpicture}
  [scale=.8,auto,every node/.style={circle,fill=blue!20},minimum size=20pt,draw=green,thick]
  \node (n2) at (1,3) {};
  \node (n3) at (1,1)  {};
  \node (n1) at (2,2)  {1};
  \node (n4) at (3,3)  {};
  \node (n5) at (3,1)  {};
  
  \node (n6)  at (6,1)  {1};
  \node (n7)  at (5,2)  {2};
  \node (n8)  at (7,3)  {3};
  \node (n9)  at (9,2)  {4};
  \node (n10) at (8,1)  {5};

  \foreach \from/\to in {n1/n2,n1/n3,n1/n4,n1/n5,n6/n7,n6/n8,n6/n9,n6/n10,n7/n8,n7/n9,n7/n10,n8/n9,n8/n10,n9/n10}
    \draw (\from) -- (\to);

\end{tikzpicture}

 \end{center} 
   \caption{\small Each node represents a link in $\mathcal{G}$. If two links cannot transmit simultaneously there is an edge between them. A minimum vertex cover is the minimum number of nodes that can cover all the edges. As it can be seen for star graph this number is 1 while in a complete graph it is the same as the numbers of the nodes 5.}
  \label{fig:vertex cover}

\end{figure}
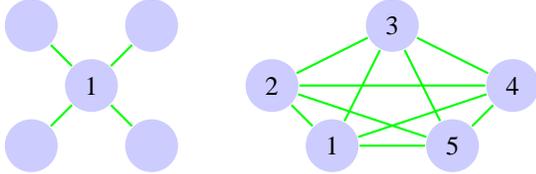

The proof is complete by understanding the graphical meaning of $\mathcal{C}$. 
To this end note the following inequality for $\mathcal{C}=1+\sum_{i}(1-x_i)\hspace{.1cm} \forall \textbf{X}$,
\begin{eqnarray}
\label{min vertex }
\begin{array}{l}
\begin{split} 
1+n-\displaystyle \max_{\textbf{X}}\sum_{i}x_i\hspace{.1cm} <\mathcal{C}
\end{split} 
\end{array}
\end{eqnarray}
where $n$ and $\displaystyle \max_{\textbf{X}}\sum_{i}x_i\hspace{.1cm}$ are respectively the number of the vertices and the maximum independent set of the interference graph $\mathcal{G}$. From graph theory it is known that the number of vertices of a graph is equal to its minimum vertex cover number plus the size of a maximum independent set (Fig.\ref{fig:vertex cover}). Therefore the right hand side of the inequality \eqref{min vertex } is the minimum vertex cover plus 1 and the minimum vertex cover of $O(\log n)$ implies $\mathcal{C}=1+\sum_{i}(1-x_i)\hspace{.1cm} \forall \textbf{X}$ to be $O(\log n)$ as well.
This completes the proof. 
\end{proof}

%However we should study the existence of $\delta'$ that can bound $\mathcal{D}(\zeta_i)-\mathcal{D}(\zeta_i+\delta_i)$ which is equivalent to the feasibility of the dual problem  of \eqref{optimation_problem}. 
%Let $\boldsymbol \delta^{*}=\underset{\boldsymbol \delta> - \boldsymbol \zeta}{\operatorname{arginf}} \hspace{.1cm} \mathcal{A}(\boldsymbol \zeta,\boldsymbol \delta) $
\section{Numerical analysis of the dependencies matrix $\mathcal{R}$}

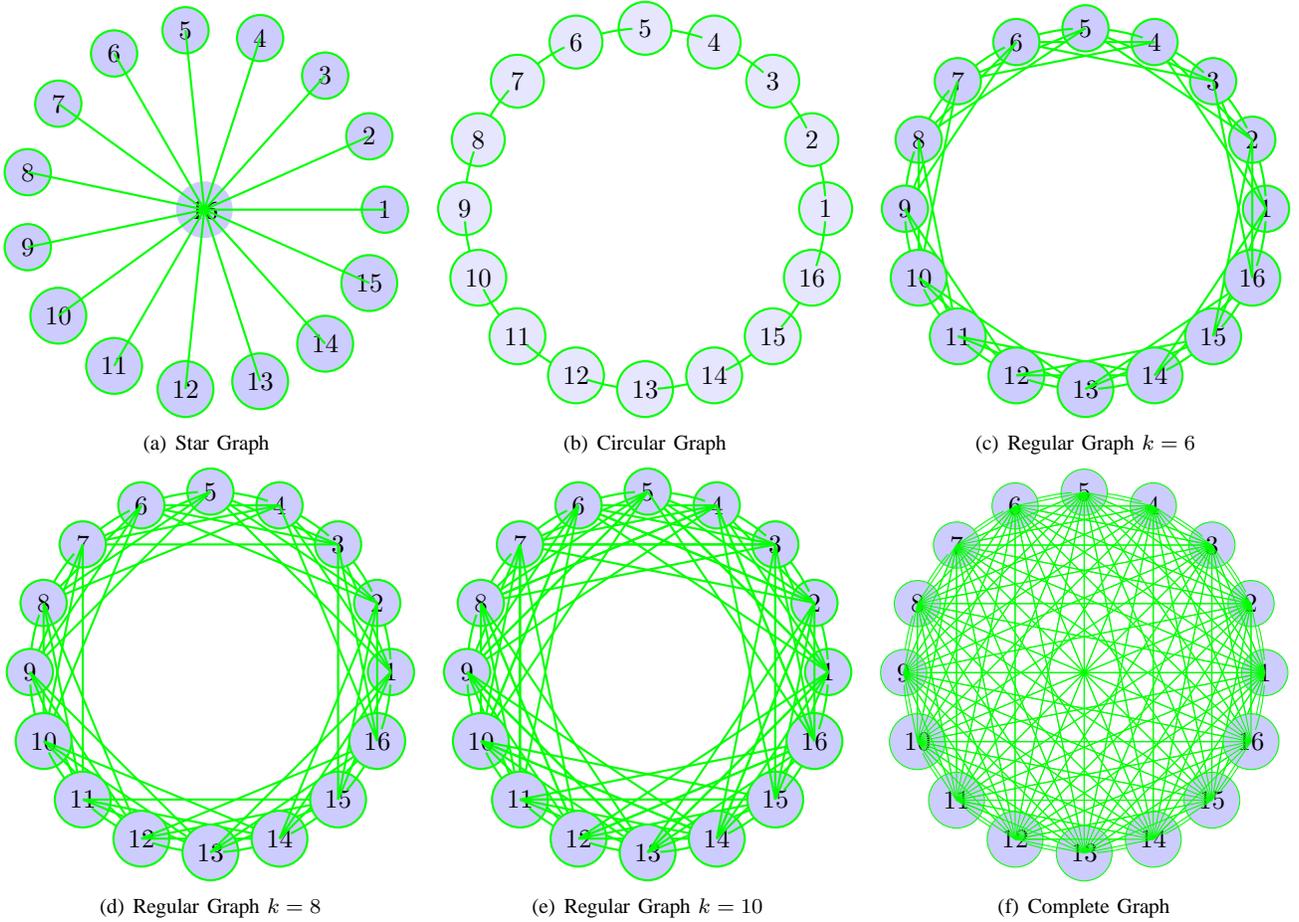
\begin{figure*}
\centering
\subfigure[Star Graph]{

\begin{tikzpicture}
% \fill[fill=blue!10!green!10!,draw=blue,dotted,thick] (0,0) circle (\r);
 [scale=.6,auto,every node/.style={circle,fill=blue!20},minimum size=10pt,draw=green,thick]

\def \n {15}
\def \radius {4cm}
\def \margin {10} % margin in angles, depends on the radius
\node (n10) at (0,0)  {16};

\foreach \s in {1,...,\n}
{
  \node[draw, circle] at ({360/\n * (\s - 1)}:\radius) {$\s$};
   \draw ({360/\n * (\s-1)}:\radius) 
 -- (0,0);

}

\end{tikzpicture}

}
\subfigure[Circular Graph] 
{

\begin{tikzpicture}
% \fill[fill=blue!10!green!10!,draw=blue,dotted,thick] (0,0) circle (\r);
 [scale=.6,auto,every node/.style={circle,fill=blue!10},minimum size=20pt,draw=green,thick]

\def \n {16}
\def \radius {4cm}
\def \margin {4} % margin in angles, depends on the radius

\foreach \s in {1,...,\n}
{
  \node[draw, circle] at ({360/\n * (\s - 1)}:\radius) {$\s$};
  \draw ({360/\n * (\s - 1)+\margin}:\radius) 
    arc ({360/\n * (\s - 1)+\margin}:{360/\n * (\s)-\margin}:\radius);   
}

\end{tikzpicture}

}
\subfigure[Regular Graph $k=6$]{
\begin{tikzpicture}
% \fill[fill=blue!10!green!10!,draw=blue,dotted,thick] (0,0) circle (\r);
 [scale=.6,auto,every node/.style={circle,fill=blue!20},minimum size=10pt,draw=green,thick]

\def \n {16}
\def \radius {4cm}
\def \margin {4} % margin in angles, depends on the radius

\foreach \s in {1,...,\n}
{
  \node[draw, circle] at ({360/\n * (\s - 1)}:\radius) {$\s$};
  \draw ({360/\n * (\s - 1)+\margin}:\radius) 
    arc ({360/\n * (\s - 1)+\margin}:{360/\n * (\s)-\margin}:\radius);
    
  \draw ({360/\n * (\s-1)}:\radius) 
 -- ({360/\n * (\s+1)}:\radius);
 
  \draw ({360/\n * (\s-1)}:\radius) 
 -- ({360/\n * (\s+2)}:\radius);

}

\end{tikzpicture}

}
\subfigure[Regular Graph $k=8$]{
\begin{tikzpicture}
% \fill[fill=blue!10!green!10!,draw=blue,dotted,thick] (0,0) circle (\r);
 [scale=.6,auto,every node/.style={circle,fill=blue!20},minimum size=10pt,draw=green,thick]

\def \n {16}
\def \radius {4cm}
\def \margin {4} % margin in angles, depends on the radius

\foreach \s in {1,...,\n}
{
  \node[draw, circle] at ({360/\n * (\s - 1)}:\radius) {$\s$};
  \draw ({360/\n * (\s - 1)+\margin}:\radius) 
    arc ({360/\n * (\s - 1)+\margin}:{360/\n * (\s)-\margin}:\radius);
    
  \draw ({360/\n * (\s-1)}:\radius) 
 -- ({360/\n * (\s+1)}:\radius);
 
  \draw ({360/\n * (\s-1)}:\radius) 
 -- ({360/\n * (\s+2)}:\radius);
 
  \draw ({360/\n * (\s-1)}:\radius) 
 -- ({360/\n * (\s+3)}:\radius);
 
}

\end{tikzpicture}

}
\subfigure[Regular Graph $k=10$]{
\begin{tikzpicture}
% \fill[fill=blue!10!green!10!,draw=blue,dotted,thick] (0,0) circle (\r);
 [scale=.6,auto,every node/.style={circle,fill=blue!20},minimum size=10pt,draw=green,thick]

\def \n {16}
\def \radius {4cm}
\def \margin {4} % margin in angles, depends on the radius

\foreach \s in {1,...,\n}
{
  \node[draw, circle] at ({360/\n * (\s - 1)}:\radius) {$\s$};
  \draw ({360/\n * (\s - 1)+\margin}:\radius) 
    arc ({360/\n * (\s - 1)+\margin}:{360/\n * (\s)-\margin}:\radius);
    
  \draw ({360/\n * (\s-1)}:\radius) 
 -- ({360/\n * (\s+1)}:\radius);
 
  \draw ({360/\n * (\s-1)}:\radius) 
 -- ({360/\n * (\s+2)}:\radius);
 
  \draw ({360/\n * (\s-1)}:\radius) 
 -- ({360/\n * (\s+3)}:\radius);
 
   \draw ({360/\n * (\s-1)}:\radius) 
 -- ({360/\n * (\s+3)}:\radius);
 
  \draw ({360/\n * (\s-1)}:\radius) 
 -- ({360/\n * (\s+4)}:\radius);
}

\end{tikzpicture}

 }
\subfigure[Complete Graph]{
\begin{tikzpicture}
% \fill[fill=blue!10!green!10!,draw=blue,dotted,thick] (0,0) circle (\r);
 [scale=.6,auto,every node/.style={circle,fill=blue!20},minimum size=5pt,draw=green]

\def \n {16}
\def \radius {4cm}
\def \margin {4} % margin in angles, depends on the radius

\foreach \s in {1,...,\n}
{
  \node[draw, circle] at ({360/\n * (\s - 1)}:\radius) {$\s$};
  \draw ({360/\n * (\s - 1)+\margin}:\radius) 
    arc ({360/\n * (\s - 1)+\margin}:{360/\n * (\s)-\margin}:\radius);
    
  \draw ({360/\n * (\s-1)}:\radius) 
 -- ({360/\n * (\s+1)}:\radius);
 
  \draw ({360/\n * (\s-1)}:\radius) 
 -- ({360/\n * (\s+2)}:\radius);
 
  \draw ({360/\n * (\s-1)}:\radius) 
 -- ({360/\n * (\s+3)}:\radius);
 
  \draw ({360/\n * (\s-1)}:\radius) 
 -- ({360/\n * (\s+4)}:\radius);
 
  \draw ({360/\n * (\s-1)}:\radius) 
 -- ({360/\n * (\s+5)}:\radius);
 
  \draw ({360/\n * (\s-1)}:\radius) 
 -- ({360/\n * (\s+6)}:\radius);
 
  \draw ({360/\n * (\s-1)}:\radius) 
 -- ({360/\n * (\s+7)}:\radius);
 
  \draw ({360/\n * (\s-1)}:\radius) 
 -- ({360/\n * (\s+8)}:\radius);
 
  \draw ({360/\n * (\s-1)}:\radius) 
 -- ({360/\n * (\s+9)}:\radius);
 
  \draw ({360/\n * (\s-1)}:\radius) 
 -- ({360/\n * (\s+10)}:\radius);
 
  \draw ({360/\n * (\s-1)}:\radius) 
 -- ({360/\n * (\s+11)}:\radius);
 
  \draw ({360/\n * (\s-1)}:\radius) 
 -- ({360/\n * (\s+12)}:\radius);
 
  \draw ({360/\n * (\s-1)}:\radius) 
 -- ({360/\n * (\s+13)}:\radius);
 
  \draw ({360/\n * (\s-1)}:\radius) 
 -- ({360/\n * (\s+14)}:\radius);
}

%\draw ({360/\n * 2}:\radius) 
% -- ({360/\n * 14}:\radius);

\end{tikzpicture}
}
\caption{\small Different graphs $\mathcal{G}$ that are used in the study of dependencies matrix $\mathcal{R}$}
  \label{fig:Graphs}
\end{figure*}

\begin{figure}
  \begin{center}
    \includegraphics[width=3.8in]{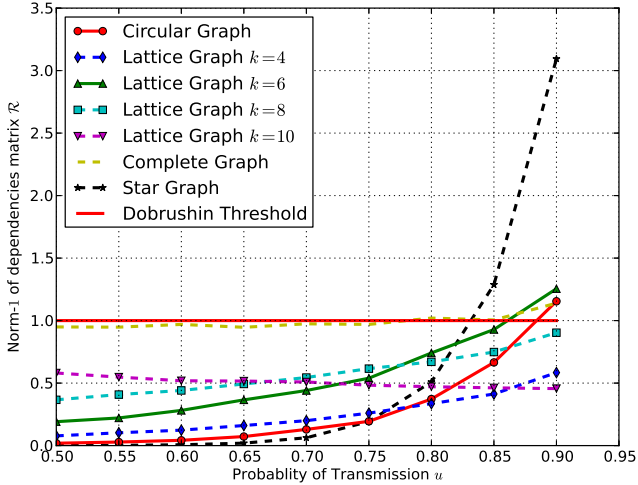}
  \end{center} 
   \caption{\small Norm-1 of the dependencies matrix $\mathcal{R}$ for different interference graphs:A service/arrival rate agnostic approach.}
  \label{fig:simulation 2}
\end{figure}

We have simulated the described CSMA. Under these simulations, the dependencies matrixes $\mathcal{R}$ for different interference graphs of Fig.\ref{fig:Graphs} are derived. We have simulated the described CSMA with the number of the links $n=16$. The probability distances and the corresponding dependencies matrix $\mathcal{R}$ are achieved by running the the simulations for $10^6$ iterations for three different scenarios:

In the first scenario it is assumed that all links use the same transmission strategy $0 < \mathcal{U} < 1$ and that they have packets to transmit all the time. This way the strategy of transmission is independent of the arrival rate $\nu_i$ and service rate $s_i$. The results are shown in Fig.\ref{fig:simulation 2}. It may be noticed that except the star graph, the norm-1 of the other dependencies matrices fall below the threshold ${\lVert \mathcal{R} \rVert}_1 =1$. This may be justified by considering the Theorem 5 and examining the minimum vertex size of the star graph with $\mathcal{C}=1$ and the rest of the graphs with $\mathcal{C}>\log(n)$. We have observed some values of more than 1 for the complete graph and circular graph but this should be due to our small numbers of the links $n$ and limited numbers of the iterations. For ${\lVert \mathcal{R} \rVert}_1 \geq 1$, each link has non-negligible effect on every other link in the system. Therefore the whole scheduling system is more than just the aggregates of distributed scheduling links. This demonstrates the emergence property of complex physical systems

\begin{figure}
  \begin{center}
    \includegraphics[width=3.8in]{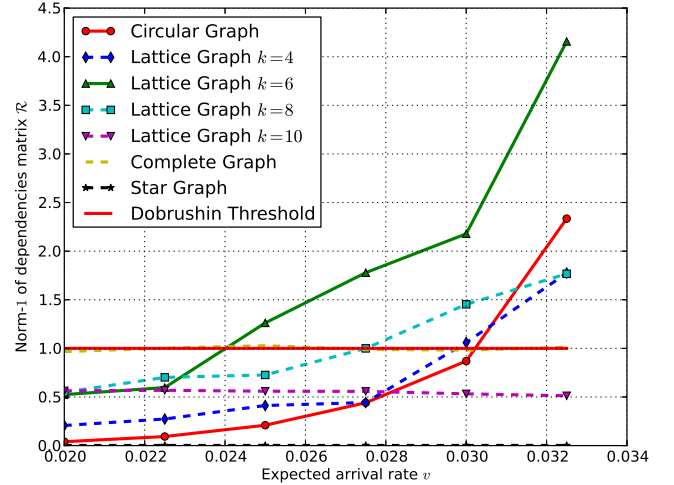}
  \end{center} 
   \caption{\small Norm-1 of the dependencies matrix $\mathcal{R}$ for different expected arrival vector of the $\boldsymbol \nu:(\nu_{i}=v), \forall i$ and different interference graphs generated using the distributed optimization of \eqref{alg1} with constant $\alpha=0.01$. }
  \label{fig:simulation 3}
\end{figure}

Another scenario is simulated where the packets arrive at the links according to Bernouli distribution of parameter $\nu$ in the range showed in Fig. \ref{fig:simulation 3}. The range carefully selected to keep it within the capacity region of the network eq. \eqref{capdef}. The links update their transmission parameters according to the \eqref{alg1} with the learning rate $\alpha=0.01$. Also the initial transmission probability is 0.5 for all links. This simulation shows that that efficiency of distributed optimization of \eqref{alg1} greatly deteriorates for the interference graphs with high $\mathcal{C}=O(\log n)$.  
Values of norm-1 ${\lVert \mathcal{R} \rVert}_1 \geq 1$ imply high coupling among distributed link optimizations where the whole system optimization cannot be decomposed to distributed link optimizations. This shows the emergence property of complex adaptive systems. That is the local observation of links under the distributed optimizations ($\textbf{s}=(s_i)$) are tied to each other to a level that the best strategy is a service-agnostic one as is predicted by Theorem 5. Note that unlike the graphs with high minimum vertex cover size of $\mathcal{C}=\log n$ (or higher), the norm-1 of star configuration stays at almost zero.

\begin{figure}
  \begin{center}
    \includegraphics[width=3.8in]{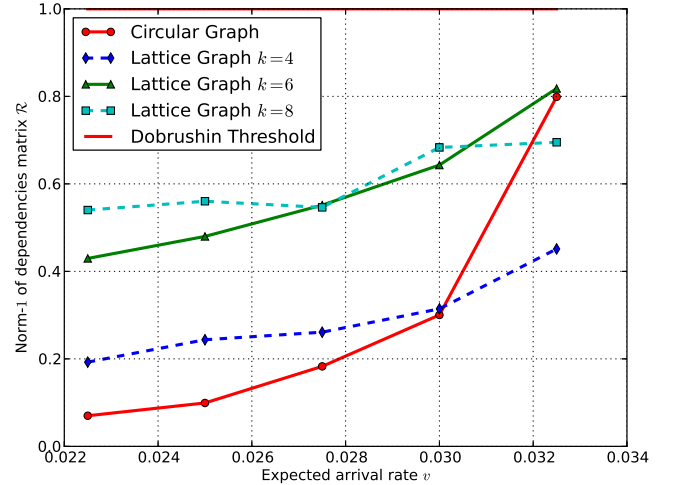}
  \end{center} 
   \caption{\small Norm-1 of the dependencies matrix $\mathcal{R}$ for different expected arrival vector of the $\boldsymbol \nu:(\nu_{i}=v), \forall i$ and different interference graphs generated using the distributed optimization of \eqref{alg1} with time variant $\alpha(t)=\frac{1}{(1+{t}^{0.3})\log(1+{t}^{0.3})}$. }
  \label{fig:simulation 4}
\end{figure}

The third scenario in Fig. \ref{fig:simulation 4} is the same as the second one with the difference that the learning rate is selected to be the time dependent according to $\alpha(t)=\frac{1}{(1+{t}^{0.3})\log(1+{t}^{0.3})}$ where $t$ is the time of the current strategy update. This time dependent learning rate has been proven to avoid complexity of the system [3]. Another strategy update mechanism is to bind the transmission probability by $\frac{1}{d_i}$ where $d_i$ is the degree of link $i$. However in the formulation of [3], the complexity avoidance concern (fast mixing condition in their context) is not part of the optimization problem and clearly not an optimal answer. For example in the case of a $k$-regular graph, the transmission strategy of the links should be less than $\frac{1}{k}$ however using the result of Fig. \ref{fig:simulation 2} it can be seen that  for no value of probability transmission the complexity emerges (The norm of the dependencies matrix stays below 1 for all transmission probabilities). The complexity concern in our paper is an internal part of the optimization formulation.

%\noindent The spectral norms, the $\gamma$-entropy, are unitary invariant and so can be expressed in terms of the singular values of $\mathcal{R}$. Assume $\mathcal{R}$ to be a complete matrix which considering $\mathcal{R}$ to have real entries that is $\mathcal{R}\mathcal{R}^{T}=\mathcal{R}^{T}\mathcal{R}$ we can write $\mathcal{I}_\gamma(\mathcal{R})=\begin{cases} -\sum_l{\gamma}^2\log (1-(\frac{\lambda_{\mathcal{R}}^l}{\gamma})^2) & \text{if} \hspace{.1cm} \lambda_{\mathcal{R}}^1<\gamma \\ \infty & \text{if} \hspace{.1cm}\lambda_{\mathcal{R}}^1 \geq \gamma.  \\ \end{cases}$ where $\lambda_{\mathcal{R}}^1$ is the principle eigenvalue of the matrix. 
%\textbf{Definition 2} Define $\mathcal{E}$ to be $\log(\sqrt{\sum_{i,j}\mathcal{R}_{ij}^2})$ as the capacity metric which can be written as $\log(\lVert \mathcal{R} \rVert_{F})$ where $\lVert \mathcal{R} \rVert_{F}$ is the Frobenius norm.
%This way term $\mathcal{E}$ provides a good valuation of the throughput.
%To understand the Frobenius norm consider the influence matrix as the adjacency matrix then the the Frobenius norm will return the 2 times the number of edge which is a normalization in terms of the existing edges in the graph. 
 
\section{Conclusion}

Cognitive services in wireless networks have provided alternative approaches for exploiting the existing resources. These services have been realized by providing the learning ability for the network elements to learn from the radio environments and their interactions with the rest of the network. 
These cognitive abilities shift the underlying models of communication system from complex physical systems to complex adaptive systems. This is because of the ability of cognitive nodes to interact with each other in a distributed way, where each node not only learn from the radio environment but also interact with other nodes. This increases the complexity of the wireless networks.

The question we address here is how arrival rate, interference graph and the simple gradient methods for adjusting the transmission rates, affect the complexity of maximum throughput optimization? We answered this question by introducing a dependencies matrix into the maximal throughput optimization in the CSMA scheduling. Beside studying the complexity of the CSMA, a direct result of our work in Theorem 5 is to prove that when the minimum vertex cover of the interference graph is logarithmic in terms of the size of vertexes $O(\log n)$, the suitable strategy for solving the optimization problem is to transmit with probability 1, i.e a service-rate agnostic approach. 

The CSMA scenario was simulated to derive the dependencies matrix and its connection with interference graph, transmission strategy and the arrival rate of the links. The result of Theorem 5 was confirmed using the result of our simulations.

\appendices
\section{Proof of Theorem 3}

\begin{proof}
We prove that the dual of \eqref{dual} is \eqref{prime}.
The Lagrangian for \eqref{dual} is
\begin{eqnarray}\label{Lagrangian}
\begin{array}{l}
\begin{split} 
\mathcal{L}(\textbf{p},\textbf{r},\boldsymbol \alpha,\gamma)=
-H(\textbf{p})+
\sum_i r_i(\sum_{\textbf{X}}-p(\textbf{X})x_i+\nu_i)+\\
\sum_{\textbf{X}}\alpha_{X}p(\textbf{X})+
\gamma(\sum_{\textbf{X}}p(\textbf{X})-1)
\end{split}
\end{array}
\end{eqnarray}
where $\mathcal{H}(\textbf{p})=-\displaystyle \sum_{\textbf{X}\in \Omega}p(\textbf{X})\log p(\textbf{X})$.
In order to derive the dual Lagrangian let's take the first derivative with respect to $p(\textbf{X})$ for all states $\textbf{X}$. This yields:
\begin{eqnarray}\label{derivitive1}
\begin{array}{l}
\begin{split} 
\partial \mathcal{L}/\partial p(\textbf{X})=\log p(\textbf{X})+1-\sum_i r_i x_i -\alpha_{\textbf{X}}+\gamma
\end{split} 
\end{array}
\end{eqnarray}
Let $\textbf{p}^{*}=\arg \displaystyle \inf_\textbf{p} \mathcal{L}$, then
\begin{eqnarray}\label{derivitive1ans}
\begin{array}{l}
\begin{split} 
p^{*}(\textbf{X})=\exp(-\gamma-1+\sum_i r_i x_i+ \alpha_{\textbf{X}})
\end{split}
\end{array}
\end{eqnarray}
This yields the dual Lagrangian function of
\begin{eqnarray}\label{dual_Lagrangian}
\begin{array}{l}
\begin{split} 
f(\textbf{r},\boldsymbol \alpha,\gamma)=\displaystyle \inf_{\textbf{p}} \mathcal{L}(\textbf{p},\textbf{r},\boldsymbol \alpha,\gamma)=
\\
-\sum_{\textbf{X}}p^{*}(\textbf{X})-\gamma+ \sum_i \nu_i r_i 
\end{split} 
\end{array}
\end{eqnarray}
To optimize the Lagrange dual function let's take the derivative with respect to $\gamma$
that is
\begin{eqnarray}\label{derivitive2}
\begin{array}{l}
\begin{split} 
\partial f/\partial \gamma=\sum_{\textbf{X}} \exp(-\gamma-1+\sum_i r_i x_i+ \alpha_{\textbf{X}})-1
\end{split} 
\end{array}
\end{eqnarray}
Setting \eqref{derivitive2} yields
\begin{eqnarray}\label{gammma}
\begin{array}{l}
\begin{split} 
\gamma^{*}=\arg \sup_{\boldsymbol \gamma} f(\textbf{r},\boldsymbol \alpha, \gamma)
\end{split} 
\end{array}
\end{eqnarray}
 that 
 \begin{eqnarray}
\label{?}
\begin{array}{l}
\begin{split} 
\exp(\gamma^{*}+1)=\sum_{\textbf{X}} \exp(\sum_i r_i x_i+ \alpha_{\textbf{X}})
\end{split}
\end{array}
\end{eqnarray}
then we have
\begin{eqnarray}
\label{gammmma2}
\begin{array}{l}
\begin{split} 
f(\textbf{r}, \boldsymbol \alpha) \equiv f(\textbf{r}, \boldsymbol \alpha, \gamma^{*})=\ln \mathcal{Z}(\textbf{r}, \boldsymbol \alpha) +\sum_i \nu_i r_i
\end{split} 
\end{array}
\end{eqnarray}
Since for all $\textbf{X}$, $\partial f/\partial \alpha_{\textbf{X}}>0 $, we have
$\boldsymbol \alpha^{*}=\textbf{0}$ where $\boldsymbol \alpha^{*}\equiv \arg \sup_{\boldsymbol \alpha} f(\textbf{r}, \boldsymbol \alpha)$
Therefore we have
\begin{eqnarray}
\label{dualOptimumFinal}
\begin{array}{l}
\begin{split} 
f(\textbf{r}) \equiv f(\textbf{r}, \boldsymbol \alpha^{*})=
-\ln \mathcal{Z}+\sum_{i} \nu_i r_i
\end{split} 
\end{array}
\end{eqnarray}
where
\begin{eqnarray}
\label{partition function}
\begin{array}{l}
\begin{split} 
\mathcal{Z}(\textbf{r}) \equiv \mathcal{Z}(\textbf{r}, \boldsymbol \alpha^{*})= \sum_{\textbf{X}} \exp(\sum_i x_ir_i)
\end{split} 
\end{array}
\end{eqnarray}
\end{proof}

\end{document}